\documentstyle[12pt,fleqn]{article}

\title{Essay on the Gamma Ray Laser\\
September 1979\footnote{Typed in 1999 after the original September 1979
manuscript}}

\author{Silviu Olariu
\thanks{Present address: Institute of Physics and Nuclear Engineering,
Department of Fundamental Experimental Physics, 
76900 Magurele, P.O. Box MG-6, Bucharest, Romania;
e-mail: olariu@ifin.nipne.ro}}

\begin{document}
\date{}
\maketitle

The coherent, high-intensity sources of light, known as lasers, represented a
great progress in modern optics. Whenever a sufficiently high population can be
stored in a higher state of a quantal system which also has a lower, relatively
unpopulated, energy level, the process of stimulated emission confers to that
source the character of a laser. We say "character", because strictly speaking,
an essential component of a laser is the resonant cavity. But if the particles
in higher state are so dense and the cross section of the process so big as to
give rise to large gains, the upper state is depopulated by a travelling wave:
we have to do in this case not with a true laser, but rather with a
superradiant source. There is a property of the laser radiation which is
conserved, even with superradiant sources: the high intensity of the radiation;
to a certain extent, it is also conserved the directionality of the radiation.

Usually the optical cavity consists of two mirrors with good reflectivity at
the operating frequency of the laser. The property of reflection is related at
optical frequencies to the interaction of the electromagnetic field with the
free electrons characteristic to the internal structure of the mirror.
Reflective materials can be found which cover the whole range of optical
frequencies. 

The basic concepts related to the optical laser, like discrete energy levels,
populating of the higher states, stimulated emission, inversion and gain, are
in principle applicable to any quantal system. Practically, it is not always
easy to accomodate the requirements for inversion and gain.

Lasers are built which make use of atomic and molecular transitions, which are 
operating in the infrared, in the visible, in the ultraviolet. The upper limit
of the photon energy attained with the excimer lasers is of about 10 eV, or 10
nm. Attempts to obtain laser radiation in the X-ray region from atomic
transitions to the inner electronic shells have failed because of the extremely
short life-times of the corresponding excited states.

The level of structure of the matter, next to the molecular and atomic one, is
represented by the atomic nucleus. A given nucleus can exist in the ground
state, or in one of its excited states. The transitions between these states
are accompanied by the emission or absorption of electromagnetic radiation. The
radiations characteristic to nuclear transitions are called gamma rays and
their energy ranges from keV's to tens of MeV.

The life-time of the nuclear excited states cover an extraordinarily wide range
of values: states with a life-time of $10^{-18}$ sec, and excited states with
life-time of years are known; a frequent life-time of a nuclear level in the 1
MeV region is 1 nsec. 

Because the nucleus is relatively isolated of the other surrounding nuclei by 
the electronic shell, the width of the levels is not affected by collisional
processes. Therefore, the cross section for many resonant transitions is the
Breit-Wigner one, which is proportional to the square of the wavelength of the
electromagnetic radiation, and lower energy gamma rays have so large cross
sections as $10^{-18}$ cm$^2$. Another consequence of the screening between the
nuclei, mediated by the electron cloud, is the existence of long-lived excited
states, called isomeric states, which correspond to highly forbidden gamma ray
transitions. 

A nucleus may be found in its excited state as a result of the decay of another
unstable nucleus and a sample of such unstable nuclei represents the common
gamma ray source. The unstable nuclei are obtained from other, stable nuclei by
irradiation with fluxes of particles, which are most conveniently neutrons.
Sometimes, the absorption by a nucleus of another particle, like a neutron,
leaves the resulting nucleus in one of its excited states, and represents a
method {\it in situ} for having nuclei in excited states.

The thermal motion of the nuclei produces a Doppler broadening of the lines,
and the energy shift known as the recoil shift, due to the momentum carried
away by the nucleus in the process of transition, often puts off resonance the
process of transition. These two effects, which are proportional to the energy,
and to the square of the transition energy, become important with gamma ray
transitions, and produce a diminishing of the transition cross section by many
orders of magnitude. Yet for transition energy not higher than 100 keV, the
binding of the nuclei in crystalline lattices eliminates both the Doppler
effect and the recoil energy shift, through what is known as the M\"{o}ssbauer
effect. 

Since a typical solid-state particle density is some $10^{22}$/cm$^3$, since 
there are means of populating higher nuclear states, and since the cross
section for the induced emission can be, under favorable conditions, some
$10^{-18}$ cm$^2$, there are no principle constraints to the construction of a
device operating as a laser at gamma-ray frequencies. Such a device would be a
superradiating gamma ray source, rather than a laser, because there is not
possible to construct the analogue of the optical cavity at gamma ray
frequencies. The purpose of this paper is to investigate the conditions for the
amplification of the gamma rays.

Despite this apparently favorable situation, the attempts to obtain stimulated
gamma radiation failed. If one tries to populate the excited states by the
conventional method of the decay of unstable nuclei, the particle density in
the excited state is some $10^{22}$/cm$^3$, corresponding to a sample
consisting only of the unstable species, times the ratio of the life-time of 
the level to the half-life of the unstable nucleus. Even in the very favorable
situation of a level of 10 $\mu$sec and a half-life of $10^3$ seconds, the
particle density in the excited state cannot exceed $10^{15}$/cm$^3$, and the
resulting gain, $10^{-3}$/cm is not sufficient for amplification. Levels with
longer life-time cannot be used, because they are broadened by imperfect sample
preparation and that diminishes the cross section, and unstable nuclei with
halflives lower than $10^3$ sec are difficult to concentrate up to
$10^{22}$/cm$^3$. 

Or one could try to populate the upper level by irradiating the sample with 
neutrons. The fraction of the nuclei in the excited state is the number of
neutrons per cm$^2$ sec times the neutron capture cross section times the
life-time of the level. \footnote{It is implicit that the excitation by 
absorption of neutrons creates a chemically different nucleus which is in 
the excited state, and the same at the decay of an unstable nucleus.} The
highest today-available neutron flux is some $10^{15}$/cm$^2$ sec, and assuming
$10^{-22}$ cm $^2$ (100 barns) for the neutron capture cross section and
$10^{-5}$ sec  for the life-time of the level, the fraction becomes $10^{-12}$,
which corresponds to an upper-state particle density not exceeding
$10^{11}$/cm$^3$, and a gain of $10^{-7}$.

The case of a neutron pulse appears more favorable: the fraction of the nuclei
in the upper, excited state is the product of the number of neutrons per cm$^2$
times the neutron capture cross section. For a neutron capture cross section of
100 barns, the neutron density resulting in a gain of 1/cm is $10^{18}$/cm$^2$,
or at least $10^{20}$ neutrons per $10^{-5}$ sec pulse duration, and that is
not a laboratory experiment.

A gain of $10^{-3}$/cm is not, strictly speaking, without significance, but the
samples appropriate for the assumed very narrow lines are probably sensitive to
the high irradiation level required to obtain the population of the upper
state. 

The previous estimations show that states with life-times below 10 $\mu$sec do
not allow enough populating to obtain relevant gains. High particle
concentrations could be obtained in the isomeric states. For long-lived states,
one can think, in principle, of samples consisting only of isomeric nuclei,
that is of densities of some $10^{22}$/cm$^3$. The difficulty here is that the
cross section of the transition to a lower state is the Breit-Wigner,
$10^{-18}$ cm$^2$, cross section, reduced by the ratio of the life-time of the
lower level to the half-life of the isomeric state, and, as before, imperfect
sample preparation and the concentration make that this ratio be something like
$10^{-5}/10^3$, and that results in a gain not exceeding $10^{-3}$/cm, under
most favorable conditions. One could think of using the transition between the
upper, isomeric state, and the lower, ground state, but the imperfect sample
preparation reduces the Breit-Wigner cross section in the ratio of the
effective to the natural life-time of the isomeric state. We are confrunted 
here with the puzzling problem of having a large concentration of nuclei in the
upper, isomeric state, which seemingly cannot be transferred to the lower
state. 

We have sometimes thought that a two photon transition could provide a way to
avoid the small matrix elements connecting the isomeric state with the lower
states. The two-photon nuclear transitions mediated by the magnetic sublevels,
which generally have cross sections up to $10^{-20}$ cm$^2$, are not
appropriate to this situation because their cross section contains the same
small matrix element, which in fact defines the existence of the isomeric 
state. Other types of multiphoton processes are clearly not useful, because 
their cross section is definitely lower than $10^{-23}$ cm$^2$.

But one possibility remains: the existence of states nearly degenerate with the
nuclear isomeric state. Roughly speaking, if the nuclei in the sample are
initially in the isomeric state, an electromagnetic field irradiating the
sample, of frequency resonant to the transition energy between the isomeric and
the nearly degenerate state, will populate the nearly degenerate state; 
the larger the cross section for single-photon, resonant transition from the
isomeric to the nearly degenerate, states, and the larger the power density in
the electromagnetic field, the higher the particle concentration in the state
degenerate with the isomeric state. The transition from the level nearly
degenerate with the isomeric state, which level will be called the normal upper
level, to a lower state has a Breit-Wigner cross section, provided the
life-times of the normal and of the lower levels are not too long. This will be
also a M\"{o}ssbauer transition, because it is the energy transferred between
the nucleus and the field which determines the M\"{o}ssbauer character of the
process, and not the absolute location of the levels. Now, if the gain
resulting from the particle concentration in the normal upper state is large
enough, a wave travelling in the active sample will be amplified and the upper
level will be depleted in the creation of a gamma ray pulse.

The size of the sample is limited by its absolute radioactivity. A 1 Ci sample
consisting of nuclei in the isomeric state of half-life $10^6$ sec 
($\approx 10$ days) would contain about $3\times 10^{16}$ isomeric nuclei. 
Assuming complete conversion of the isomeric nuclei to the lower state, the 
gamma-ray pulse would carry an energy of about 50 Joules. We also note that a 
global recoil momentum becomes important due to the high concentrations in the 
sample.

The many intersections which occur in the Nilsson diagrams suggest that nearly
degenerate states have a real existence. The fact that one of the states has to
be isomeric, and the selection rule requirements, represent, of course,
additional constraints. In any case, two nearly degenerate states with an
energy separation corresponding to optical frequencies are beyond the resolving
power of the existing gamma-ray spectroscopy. Spontaneous transitions between
such states are also imperceptible because that probability is proportional to
the third power of the frequency of transition.

The cross section of a single photon radiative exchange between the states
$n,n_0$ of a quantal system and the mode of frequency $\omega$ and width
$\gamma$ of the electromagnetic field is, apart from numerical coefficients,
\begin{equation}
\sigma\sim \frac{e^2}{\hbar c}\frac{\omega}{\gamma}|v_{nn_0}|^2 ,
\end{equation}
where $ev$ is the reduced matric element $v_\alpha$ defined in Eq. 44 of the
Third Gamma Optical Paper.\footnote{Note added in May 1999: the referenced
work is S. Olariu et al., Phys. Rev. C {\bf 23}, 50 (1981); it was submitted 
for publication on 30 November 1979.} Since the natural width of
a level is proportional to $\frac{e^2}{\hbar c^3}\omega^3 |v_{nn_0}|^2,$ and if
the width of the mode is the same as the width of the level, then substituting
the above expression in the expression, Eq. 1, for $\sigma$ gives the
$\lambda^2$ dependence of the Breit-Wigner cross section. 

In our case, the width $\gamma$ of the normal upper state is determined by the
transition to the lower state (see Figure). The width is roughly
\begin{equation}
\gamma\sim \frac{e^2}{\hbar c^3}\omega_{nl}^3 |v_{nl}|^2 .
\end{equation}
The cross section for a single photon transition between the isomeric and the
normal upper state is, according to Eq. 1,
\begin{equation}
\sigma_{ni}\sim \frac{e^2}{\hbar c}\frac{\omega_{ni}}{\gamma}|v_{ni}|^2 .
\end{equation}
Substituting in Eq. 3 the expression of the width of the normal upper state,
Eq. 2, gives for the cross section $\sigma_{ni}$,
\begin{equation}
\sigma_{ni}\sim
\left(\frac{c}{\omega_{nl}}\right)^2\frac{\omega_{ni}}{\omega_{nl}}
\frac{|v_{ni}|^2}{|v_{nl}|^2} .
\end{equation}
The first factor represents the Breit-Wigner cross section between the states
$n$ and $l$ and Eq. 4 states that the cross section for a single photon
transition between the nearly degenerate states, $n$ and $i$, is the typical
$10^{-18}$ cm$^2$, Breit-Wigner cross section, reduced in the ratio of the
energy separation to the energy of transition and by a factor depending on he
type of transitions between the states $ni$ and $nl$. For example, the
transition $i\leftrightarrow l$ could be E3 and the transitions
$n\leftrightarrow i, n\leftrightarrow l$ be E1 and E2; depending on which of
them is the E1, the ratio of the matrix elements is larger or smaller than 1.
It is, therefore, not possible to obtain an {\it a priori} estimate of the
cross section $\sigma_{ni}$.

If the sample consisting of nuclei in the isomeric state $i$ is illuminated by
a pulse of optical radiation, the probability of having the nucleus in the
normal upper state $n$ is the number, $N_{/cm^2}$, of optical photons per
cm$^2$ which crosses the sample multiplied by the transition cross section
$\sigma_{ni}$, provided the duration of the pulse is shorter or comparable to
the width of the state $n$. In order to obtain the significant gamma ray gain
of 1/cm with a gamma-ray cross section of $10^{-18}$ cm$^2$ we have to have a
particle density in the state $n$ of $10^{18}$ nuclei/cm$^3$. That means that
the product $N_{/cm^2}\sigma_{ni}$ has to be about $10^{-4}$. We have explained
previously that the radioactivity from the isomeric state limits the number of
the nuclei in the sample to about $10^{18}$ nuclei. If the length of the sample
is 1 cm, its transverse area will be about $10^{-3}$ cm$^2$.

A pulse of 10 Joule of photon energy 1 eV corresponds to about $10^{20}$
photons, or $10^{23}$ photons/cm$^2$ crossing the sample per pulse. That means
that a cross section $\sigma_{ni}$ of about $10^{-27}$ cm$^2$ would result in
the significant gain of 1/cm. From Eq. 4, we see that the ratio
$\omega_{ni}/\omega_{nl}$ is about $10^{-4}$, while the ratio of the matrix
elements may be small as well as large. It is therefore not unreasonable to
expect cross sections $\sigma_{ni}$ in the range $10^{-24}$ cm$^2$.

The fine investigation of the nuclear structure in the vicinity of a given
state has, of course, a fundamental importance. We have argued here that, if a
state nearly degenerate with long-lived isomeric states exists, which is
related to this and to the lower states by appropriate selection rules, this
occurrence will open a way toward the gamma ray laser.

The high-resolution investigation of the nuclear structure near an isomeric
state is a surprisingly simple problem. The idea is that, if the sample
consisting of isomeric nuclei is immersed in tunable optical radiation, the
resonance of the optical radiation to the sought for nearly degenerate
transition energy would result in the increase of the radioactivity of the
sample. 

That will happen whenever the product $N_{/cm^2}\sigma_{ni}/\tau_n$ which
represents the probability of the emission of gamma rays through the process
$i\leftrightarrow n\leftrightarrow l$ is larger than the probability of
spontaneous emission from the state $i$, $1/\tau_i$; $\tau_n$ and $\tau_i$ are
the life-times of the states $n$ and $i$, respectively:
\begin{equation}
N_{/cm^2}\sigma_{ni}>\tau_n/\tau_i .
\end{equation}
Values of $\tau_n$ of practical interest are $\tau_n< 10 \mu$sec (because of
the difficulties in the preparation of the samples) and for $\tau_i$ we assume
$10^6$ seconds (10 days). Because we need radiation, tunable over a wide range
of frequencies, it seems appropriate to take $N_{/cm^2}\approx 10^{18}$/cm$^2$
per pulse. Then the lower limit of measurable $\sigma_{ni}$ is of about
$10^{-30}$ cm$^2$. Of course, this technique of measurement of $\sigma_{ni}$
requires the comparing of the signals obtained with the optical field on, and
off. Weak gamma ray sources (1 mCi) and very small particle concentrations of
isomeric nuclei can be used. This method is different from the radiofrequency
resonance techniques, because it is not the absorption of the optical power
which matters, but its effect on the large, gamma ray, transition. The nuclides
with isomeric states of half-life longer than 10 days are listed further:

\vspace*{0.5cm}

half-life $>$ 10 d\\
Nb$^{92}$, Nb$^{93}$, Tc$^{95}$, Tc$^{97}$, Rh$^{102}$, Ag$^{108}$, 
Ag$^{110}$, Cd$^{115}$, Sn$^{117}$, Sn$^{119}$, Sn$^{121}$, Te$^{121}$, 
Te$^{123}$, Te$^{125}$, Te$^{127}$, Te$^{129}$, Xe$^{131}$, Pm$^{148}$, 
Ho$^{166}$, Lu$^{174}$, Lu$^{177}$, Hf$^{178}$, Hf$^{179}$, Re$^{184}$, 
Re$^{186}$, Ir$^{192}$, Ir$^{193}$, Ir$^{194}$, Bi$^{210}$, Am$^{242}$

\vspace*{0.5cm}

Since the tunability range of the optical power, multiplied by the number of
isomeric states and divided by the average separation between the nuclear
levels represents the {\it statistical} probability of finding a convenient
degenerate pair, there is a few percent chance for the existence of the pair.
But, as very little is known about the fine actual nuclear structure, a search
for degenerate states could demonstrate equally no structure as well as reveal
surprising things.

\end{document}